\documentclass[aps,prl,twocolumn,superscriptaddress,showpacs,longbibliography]{revtex4-2}
\usepackage{amsfonts,amsmath,bm}
\usepackage{mathtools}
\usepackage{graphicx}
\usepackage{epstopdf}
\usepackage[dvipsnames]{xcolor}

\DeclarePairedDelimiter\norm{\lVert}{\rVert}%
\DeclarePairedDelimiter\abs{\lvert}{\rvert}%
\makeatletter
\let\oldabs\abs
\def\abs{\@ifstar{\oldabs}{\oldabs*}}
\let\oldnorm\norm
\def\norm{\@ifstar{\oldnorm}{\oldnorm*}}
\makeatother

\begin{document}
	
	\title{Two-phonon propagation in a 4-level thermal quantum nanomachine}
	
	\author{Y. Lai}
	\affiliation{School of Physics and CRANN Institute, Trinity College Dublin, Dublin 2, D02 PN40, Ireland}
	\affiliation{The Blackett Laboratory, Department of Physics, Imperial College London, South Kensington Campus, SW7 2AZ, London, United Kingdom} 
	
	\author{C. McDwyer}
	\affiliation{School of Physics and CRANN Institute, Trinity College Dublin, Dublin 2, D02 PN40, Ireland} 
	
	\author{P. Karwat}
	\affiliation{School of Physics and CRANN Institute, Trinity College Dublin, Dublin 2, D02 PN40, Ireland} 
	\affiliation{Department of Theoretical Physics, Wroc\l{}aw University of Science and Technology, Wybrze\.ze Wyspia\'nskiego 27, 50-370 Wroc\l{}aw, Poland}

	\author{O. Hess}
	\affiliation{School of Physics and CRANN Institute, Trinity College Dublin, Dublin 2, D02 PN40, Ireland} 
	\affiliation{The Blackett Laboratory, Department of Physics, Imperial College London, South Kensington Campus, SW7 2AZ, London, United Kingdom} 
	
	\date{\today}
	
	\begin{abstract}
		Acoustic waves, as science of sound, is an established field in physics. In analogy with light, engineers adapt tools from optic fields for manipulating sound waves to fight with cancer cells. 
		Here we present a heat-gradient driven nanomachine concept for simultaneous emission of two phonons, i.e. a convertion of heat into a sonic wave output. Our theoretical work sheds light on nanoscale components or systems that could be used for future ultrasound devices.
	\end{abstract}
	

	\maketitle
	
	\section{\label{sec:introduction}Introduction} 
	Recent progress in polariton-driven phonon lasing \cite{Chafatinos:20} shows that there is much more to discover \cite{Wang:22,Keitel:21}. As lasers in optics, as sound waves in acoustics, in analogy with light they could be used as a tool, i.e. acoustic tweezer \cite{Jiang:22,Gu:20}, and play an important role to fight with cancer cells. Scientists are adopting ideas from optics to manipulate sound weaves \cite{Jooss:22,Yves:22} instead. Recent experimental works suggest that ultrasound could not only trap, but also compress a cancer cell \cite{Zeng:22}. Another motivation to study acoustics is fact, that a phononic medium offers a wider range of wavelengths, usually not accessible to conventional lasers \cite{Gu:98,Ojambati:21,McKenna:21}.
	
	In this paper, we present a thermal quantum nanomachine, i.e. a phonon lasing concept \cite{Karwat:22} based on 4-level system with two 2-level subsystems, able to convert heat into phononic output. As typical ultrasound devices used to fight with cancer cells operate in the range from 20 kHz to several GHz, we focus on energy scale with tenths of meV. In the end, we analyze some thermodynamic aspects of our phonon medium.
	
	\section{\label{sec:model}Model}
	
	\begin{figure}[t]
		\centering
		\includegraphics[width=85mm]{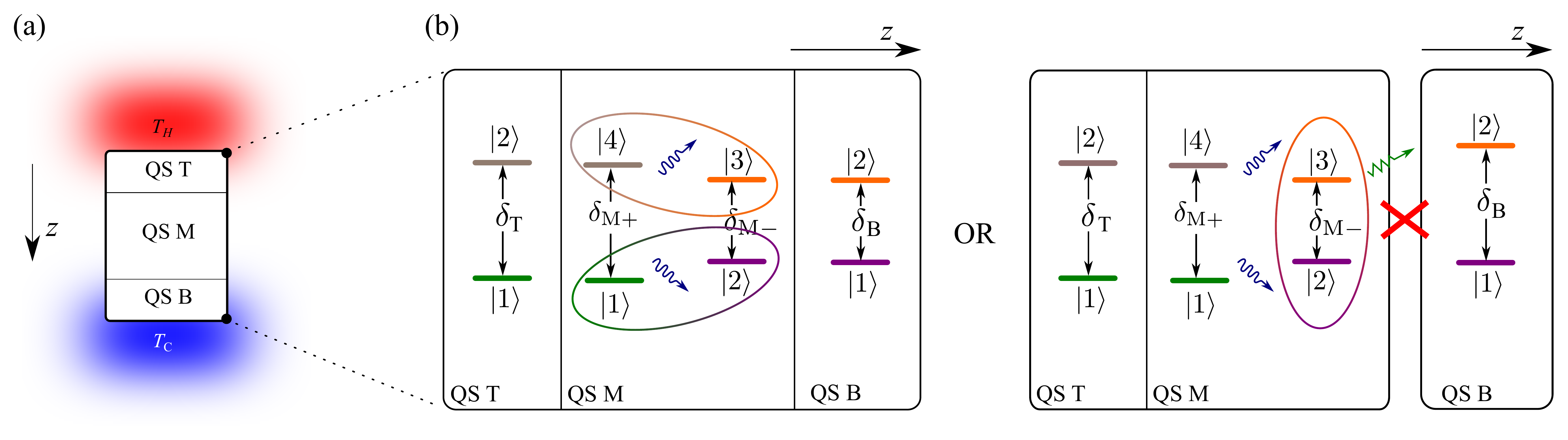}
		\caption{(a) Sketch of the system. (b) Two scenarios, \textit{left:} generation of two phonons, i.e. transitions $|4\rangle$ $\rightarrow$ $|3\rangle$ and $|2\rangle$ $\rightarrow$ $|1\rangle$ in the middle subsystem QS M.; \textit{right:} generation of a~photon, i.e. transitions $|3\rangle$ $\rightarrow$ $|2\rangle$.}
		\label{fig:4lvl}
	\end{figure}
	 The system under study illustrated in Fig.~\ref{fig:4lvl} is composed of centre 4-level quantum subsystem (QS M) interacting with two 2-level subsystems (QS L/R) at adjacent sides, while the whole system is coupled to heat baths with different temperatures. The heat-gradient \cite{chamon:11,hasegawa:18, gluza:21} is necessary for excitation flow, that allows the transitions between $|4\rangle$ $\leftrightarrow$ $|3\rangle$ and $|1\rangle$ $\leftrightarrow$ $|2\rangle$ in the middle subsystem to be accompanied with phonon emission and absorption. These two-level subsystems work as filters to avoid thermalization with a mean temperature in QS M. We assume that only the middle subsystem is coupled to external lattice displacement field.
	 
	 The QSs are coupled in a way that excitations can be exchanged between adjacent sites as denoted in Fig.~\ref{fig:4lvl}(b), i.e. the middle system interacts with both parts (i.e. top and bottom), while the interaction between top and bottom system is suppressed. The corresponding Hamiltonian reads
	 \begin{eqnarray}
	 	\label{ham_int}
	 	\hat{H}_{\mathrm{int}}= \lambda_{\mathrm{MT}}\bigg( \hat{h}^{(\mathrm{TM})}\otimes \hat{1}^{\mathrm{(B)}} \bigg) 
	 	+\lambda_{\mathrm{MB}}\bigg(\hat{1}^{\mathrm{(T)}}\otimes  \hat{h}^{(\mathrm{MB})} \bigg)
	 \end{eqnarray}
	 where $\lambda_{\mathrm{MT/MB}}$ is the coupling parameter. The coupling
	 \begin{equation}
	 	\hat{h}^{(\mathrm{TM})} = \hat{P}_{21}^{\mathrm{(T)}} \otimes \bigg( \hat{P}_{12}^{\mathrm{(M)}} + \hat{P}_{14}^{\mathrm{(M)}} + \hat{P}_{23}^{\mathrm{(M)}} + \hat{P}_{24}^{\mathrm{(M)}}\bigg) + \mathrm{h.c.}
	 \end{equation}
	 is given by the respective projection operators and analogous for $\hat{h}^{(\mathrm{MB})}$. The coupling is taken to be weak such that the energy contribution of the interaction is small compared to the energy contained in the system.
	 
	 Each of the two edge QSs is coupled locally to a heat bath of different temperature. To describe the coupling, we make use of a Quantum Master Equation within a Lindblad form, which accounts for the non-equilibrium situation in our system. For this, we set up the equation of motion for the density matrix $\hat{\rho}$ via
	 \begin{equation}
	 	\frac{d\hat{\rho}}{dt} = -\frac{i}{\hbar}[\hat{H}_{\mathrm{sys}}+\hat{H}_{\mathrm{int}},\hat{\rho}] + \hat{D}_{\mathrm{H}}(\hat{\rho}) + \hat{D}_{\mathrm{C}}(\hat{\rho})\,.
	 \end{equation}
	 $\hat{D}_{\mathrm{H}}$ and $\hat{D}_{\mathrm{C}}$ are the dissipators to the hot (left) and cold (right) heat bath, respectively,
	 \begin{equation}
	 	\hat{D}_{\mathrm{H}}(\hat{\rho}) = \sum_{k = 1}^{2}\Gamma_{k}(T_{\mathrm{H}})\bigg(\hat{L}^{(\mathrm{T})}_{k}\hat{\rho}\hat{L}^{(\mathrm{T})\dagger}_{k} 
	 	- \frac{1}{2}[\hat{L}^{(\mathrm{T})\dagger}_{k}\hat{L}^{(\mathrm{T})}_{k}, \hat{\rho}]_{+} \bigg)
	 \end{equation}
	 with the Lindblad operators 
	 \begin{equation*}
	 	\hat{L}^{(\mathrm{T})}_{1} = \hat{P}^{(\mathrm{T})}_{21}\otimes\hat{1}^{(\mathrm{M,B})},\qquad
	 	\hat{L}^{(\mathrm{T})}_{2} = \hat{P}^{(\mathrm{T})}_{12}\otimes\hat{1}^{(\mathrm{M,B})} \, .
	 \end{equation*}
	 \begin{equation}
	 	\Gamma_{k}(T_{\mathrm{H}}) = \frac{\gamma}{1+\exp \big\{(-1)^{k-1}\delta_{\mathrm{L}}/k_{\mathrm{B}}T_{\mathrm{H}}\big\}}.
	 \end{equation}
	 containing the distribution function \cite{breuer2002theory}. The dissipator describing the coupling to the cold (right) heat bath  $\hat{D}_{\mathrm{C}}(\hat{\rho})$ is analogue. \\
	 Our goal is to achieve an inversion between states: $|4\rangle ^{(\mathrm{M})} \rightarrow |3\rangle ^{(\mathrm{M})}$, $|2\rangle ^{(\mathrm{M})}~\rightarrow~|1\rangle^{(\mathrm{M})}$, which later will be coupled to a phonon mode, and $|3\rangle ^{(\mathrm{M})} \rightarrow |2\rangle ^{(\mathrm{M})}$ (see Fig.~\ref{fig:4lvl}(b)) for the coupling to a photon mode. 
	 
	 The energy of typical acoustic phonons lie in the order of a few meV. Setting the energy of the lower state $|1\rangle^{(\mathrm{M})}$ to zero, we accordingly chose $\delta_{\mathrm{M_{+}}} = 30$~meV and $\delta_{\mathrm{M_{-}}} = 25$~meV, thus $\delta_{\mathrm{M_{+}}} - \delta_{\mathrm{M_{-}}}= 5$~meV. The energies of the edge state are set to $\delta_{\mathrm{T}} = \delta_{\mathrm{M_{+}}}=30$~meV  and $\delta_{\mathrm{B}} = \delta_{\mathrm{M_{-}}}=25$~meV. If not stated otherwise, the parameters are set to $\lambda=\lambda_{\mathrm{MT}} = \lambda_{\mathrm{MB}} = 0.03$~meV, $\gamma=\gamma_{\mathrm{H}} = \gamma_{\mathrm{C}} = 3$~ps$^{-1}$. As initial condition we assume that the whole system is in its ground state $|1\rangle ^{(\mathrm{T})}\otimes |1\rangle ^{(\mathrm{M})} \otimes |1\rangle ^{(\mathrm{B})}$. 
	 
	 \section{phonon lasing}
	 For simplicity, we assume two single acoustic-phonon modes, as realized in a cavity formed, e.g., in semiconductor superlattices.
	 Denoting $b,b^{\dagger}$ as the bosonic operators of the phonon mode and $\omega$ their frequency, the phonon Hamiltonian is composed of the free phonon system and the carrier-phonon coupling
	 \begin{equation}
	 	\hat{H}_{\mathrm{ph}} = \hbar \omega_{1} \hat{b}_{1}^{\dagger}\hat{b}_{1} + \hbar \omega_{2} \hat{b}_{2}^{\dagger}\hat{b}_{2} + \hat{H}_{\mathrm{c-ph}} \,.
	 \end{equation}
	 The energy of the phonon is chosen to be resonant with the energy difference in QS~M with $\hbar\omega=5$~meV. The phonon mode is coupled only to the middle QS~M via
	 \begin{equation}
	 	\hat{H}_{\mathrm{c-ph}} = \hat{1}^{(\mathrm{T})} \otimes \hat{h}^{(\mathrm{M})} \otimes \hat{1}^{(\mathrm{B})}
	 \end{equation}
	 with
	 \begin{equation}
	 	\label{hm}
	 	\hat{h}^{(\mathrm{M})} = \hbar g\big(\hat{b}^{\dagger}\hat{P}^{(\mathrm{M})}_{34} + \hat{b}\hat{P}^{(\mathrm{M})}_{43} + \hat{b}^{\dagger}\hat{P}^{(\mathrm{M})}_{12} + \hat{b}\hat{P}^{(\mathrm{M})}_{21}\big).
	 \end{equation}
	 Here $g$ is the real coupling constant between the QS and phonon mode. We assume that the system can relax from the state $|4\rangle^{(\mathrm{M})}$ to state $|3\rangle^{(\mathrm{M})}$ and from the state $|3\rangle^{(\mathrm{M})}$ to state $|2\rangle^{(\mathrm{M})}$ by the emission of a phonon, while the reverse transition is possible by absorbing a phonon.
	 
	 A measurable quantity for the phonons is the lattice displacement, which is connected to the phonon operators via 
	 \begin{equation}
	 	\langle  \hat{u} \rangle = u_{0} \left(\langle\hat{b}^{\dagger} \rangle +\langle \hat{b}^{} \rangle \right) =  u^{(+)} + u^{(-)},
	 \end{equation}
	 where we defined $u^{(+)} = u_{0}\langle\hat{b}\rangle$ and $u^{(-)} = u_{0}\langle\hat{b}^{\dagger}\rangle$ as well as the single phonon amplitude $u_{0}$. 
	 
	 Heisenberg equation.
	 \begin{equation*}
	 	i\hbar\frac{d}{dt}\hat{b}^{\dagger} = -\hbar\omega\hat{b}^{\dagger} - \hbar g|3\rangle\langle2|
	 \end{equation*}
	 \begin{equation*}
	 	i\hbar\frac{d}{dt}\hat{b} = +\hbar\omega\hat{b} + \hbar g|2\rangle\langle 3|
	 \end{equation*}
	 \begin{equation*}
	 	\frac{d}{dt}\hat{b}^{\dagger} = i\omega\hat{b}^{\dagger} + i g|3\rangle\langle2|
	 \end{equation*}
	 \begin{equation*}
	 	\frac{d}{dt}\hat{b} = -i\omega\hat{b} - i g|2\rangle\langle 3|
	 \end{equation*}
	 \begin{equation*}
	 	\frac{d}{dt}u = i u_{0}\omega(\langle\hat{b}^{\dagger}\rangle-\langle\hat{b}\rangle) + i u_{0} g(|3\rangle\langle 2| - |2\rangle\langle 3|)
	 \end{equation*}
	 
	 Introducing the phonon coupling, we extend the equations of motion to
	 \begin{equation}
	 	\frac{d\hat{\rho}}{dt} = -\frac{i}{\hbar}[\hat{H}_{\mathrm{sys}}+\hat{H}_{\mathrm{int}}+\hat{H}_{\mathrm{ph}}  ,\hat{\rho}] + \hat{D}_{\mathrm{H}}(\hat{\rho}) + \hat{D}_{\mathrm{C}}(\hat{\rho}),
	 	\label{test}
	 \end{equation}
	 For the lattice displacement this leads to the following rate Eq. \eqref{test}.
	 \begin{equation}
	 	\frac{du_{1}}{dt} = -\Gamma u_{1} + iC \bigg(\rho_{34}^{(\mathrm{M})}(t) + \rho_{43}^{(\mathrm{M})}(t)\bigg),
	 \end{equation}
	 \begin{equation}
	 	\frac{du_2}{dt} = -\Gamma u_{2} + iC \bigg(\rho_{12}^{(\mathrm{M})}(t) + \rho_{21}^{(\mathrm{M})}(t)\bigg),
	 \end{equation}
	 where we introduced the phonon dephasing rate $\Gamma$ (assuming the same for both modes) and defined the coupling constant $C = u_{0}g$.
	 For our simulations we assume $\Gamma = 2$~ps$^{-1}$, $g = 2.25$~ps$^{-1}$, $u_{0} = 20$~pm. We further assume that always a very small, but finite displacement is present.
	
	Then the dynamics can be described by currents. We take a trace the Quantum Master Equations over all the subsystems at first. The reduced density operators only describe the states for corresponding subsystems. The diagonal elements of the reduced density matrix indicate the occupation of states in corresponding subsystems.
\begin{equation}
	\rho^\mathrm{(L)}=Tr_\mathrm{MR} (\rho),
	\rho^\mathrm{(R)}=Tr_\mathrm{LM} (\rho),
	\rho^\mathrm{(M)}=Tr_\mathrm{LR} (\rho).
	\label{}
	\end{equation}
The time evolution equations of these occupations of the energy levels in subsystems are derived from reduced master equations
\begin{equation}
	\dot{\rho}^\mathrm{(L)}=Tr_\mathrm{MR} (\dot{\rho}), 
	\dot{\rho}^\mathrm{(R)}=Tr_\mathrm{LM} (\dot{\rho}), 
	\dot{\rho}^\mathrm{(M)}=Tr_\mathrm{LR} (\dot{\rho}).
	\label{}
\end{equation}
The transitions between different states result in the change of states. To be consistent with previous research \cite{RefrigeratorbyPhotons}, we define the thermal particle currents to describe the transition induced by the thermal reservoirs
	\begin{equation}
	J^\mathrm{(L)}_{1\rightarrow2}=	\Gamma^{\mathrm{(L)}}_\mathrm{1}(T_\mathrm{H}) \rho^\mathrm{(L)}_{11}-	\Gamma^{\mathrm{(L)}}_\mathrm{2}(T_\mathrm{H})\rho^\mathrm{(L)}_{22},
	\label{}
	\end{equation}
	\begin{equation}
	J^\mathrm{(R)}_{1\rightarrow2}=\Gamma^{\mathrm{(R)}}_\mathrm{1}(T_\mathrm{C}) \rho^\mathrm{(R)}_{11}-\Gamma^{\mathrm{(R)}}_\mathrm{2}(T_\mathrm{C})\rho^\mathrm{(R)}_{22},
	\label{}
	\end{equation}
	\begin{equation}
	J^\mathrm{(M)}_{\mathrm{D},1\rightarrow2}=\Gamma^{\mathrm{(M12)}}_\mathrm{1}(T_\mathrm{sys}) \rho^\mathrm{(M)}_{11}-\Gamma^{\mathrm{(M12)}}_\mathrm{2}(T_\mathrm{sys}) \rho^\mathrm{(M)}_{22},
	\label{}
	\end{equation}
	\begin{equation}
	J^\mathrm{(M)}_{\mathrm{D},3\rightarrow4}=\Gamma^{\mathrm{(M34)}}_\mathrm{1}(T_\mathrm{sys}) \rho^\mathrm{(M)}_{33}-\Gamma^{\mathrm{(M34)}}_\mathrm{2}(T_\mathrm{sys})\rho^\mathrm{(M)}_{44}.
	\label{}
	\end{equation}
Then the independent equations of occupation (diagonal terms of density matrix) can be rewritten as
\begin{equation}
	\frac{d }{dt}\rho^\mathrm{(L/R)}_{22}=J^\mathrm{(L/R)}_{1\rightarrow2}+J^\mathrm{(M)}_{\mathrm{H/C}},
	\label{}
	\end{equation}

	\begin{equation}
	\frac{d }{dt}\rho^\mathrm{(M)}_{22}=J^\mathrm{(M)}_{3\rightarrow2}-J^\mathrm{(M)}_{2\rightarrow1},
	\label{}
	\end{equation}
	\begin{equation}
	\frac{d }{dt}\rho^\mathrm{(M)}_{33}=J^\mathrm{(M)}_{4\rightarrow3}-J^\mathrm{(M)}_{3\rightarrow2},
	\label{}
	\end{equation}
	\begin{equation}
	\frac{d }{dt}\rho^\mathrm{(M)}_{44}=J^\mathrm{(M)}_{1\rightarrow4}-J^\mathrm{(M)}_{4\rightarrow3},
	\label{}
	\end{equation}
	where
	\begin{equation}
	J^\mathrm{(M)}_{\mathrm{H}}=-J^\mathrm{(M)}_{\mathrm{H},1\rightarrow4}+J^\mathrm{(M)}_{\mathrm{H},4\rightarrow3}+J^\mathrm{(M)}_{\mathrm{H},3\rightarrow2}+J^\mathrm{(M)}_{\mathrm{H},2\rightarrow1},
	\label{}
	\end{equation}
		\begin{equation}
J^\mathrm{(M)}_{\mathrm{C}}=J^\mathrm{(M)}_{\mathrm{C},1\rightarrow4}+J^\mathrm{(M)}_{\mathrm{C},4\rightarrow3}-J^\mathrm{(M)}_{\mathrm{C},3\rightarrow2}+J^\mathrm{(M)}_{\mathrm{C},2\rightarrow1},
	\label{}
	\end{equation}

	\begin{equation}
	J^\mathrm{(M)}_{2\rightarrow1/4\rightarrow3}=\sum_{\mathrm{i}=\mathrm{H,C,Ph,D}}J^\mathrm{(M)}_{\mathrm{i},2\rightarrow1/4\rightarrow3},
	\label{}
	\end{equation}
		\begin{equation}
J^\mathrm{(M)}_{3\rightarrow2/1\rightarrow4}=J^\mathrm{(M)}_{\mathrm{H},3\rightarrow2/1\rightarrow4}+J^\mathrm{(M)}_{\mathrm{C},3\rightarrow2/1\rightarrow4},
	\label{}
	\end{equation}
 
	We assume that our phonon laser operates in steady-state ($d\rho^\mathrm{(M)}_{ii}/dt=0$), obtaining: $	J^\mathrm{(L)}_{1\rightarrow2}=-J^\mathrm{(M)}_{\mathrm{H}}$, $	J^\mathrm{(R)}_{1\rightarrow2}=-J^\mathrm{(M)}_{\mathrm{C}}$, $J^\mathrm{(M)}_{3\rightarrow2}=J^\mathrm{(M)}_{2\rightarrow1}$, $J^\mathrm{(M)}_{4\rightarrow3}=J^\mathrm{(M)}_{3\rightarrow2}$, $J^\mathrm{(M)}_{1\rightarrow4}=J^\mathrm{(M)}_{4\rightarrow3}$.

	As shown in the previous research, the energy mismatch have a large negative impact on the interaction strength \cite{Pawel}. We assume that non-resonant interaction is negligible weak, therefore the non-resonant transition will be neglected. The relations between currents become
\begin{equation}
	J^\mathrm{(L)}_{1\rightarrow2}=-J^\mathrm{(M)}_{\mathrm{H}}=J^\mathrm{(M)}_{\mathrm{H},1\rightarrow4},
	\label{}
	\end{equation}
	\begin{equation}
	J^\mathrm{(R)}_{1\rightarrow2}=-J^\mathrm{(M)}_{\mathrm{C}}=J^\mathrm{(M)}_{\mathrm{C},3\rightarrow2},
	\label{}
	\end{equation}
	\begin{equation}
	J^\mathrm{(M)}_{\mathrm{Ph},2\rightarrow1}+J^\mathrm{(M)}_{\mathrm{D},2\rightarrow1}=J^\mathrm{(M)}_{\mathrm{C},3\rightarrow2},
	\label{}
	\end{equation}
	\begin{equation}
	J^\mathrm{(M)}_{\mathrm{C},3\rightarrow2}=J^\mathrm{(M)}_{\mathrm{Ph},4\rightarrow3}+J^\mathrm{(M)}_{\mathrm{D},4\rightarrow3},
	\label{}
	\end{equation}
	\begin{equation}
	J^\mathrm{(M)}_{\mathrm{Ph},4\rightarrow3}+J^\mathrm{(M)}_{\mathrm{D},4\rightarrow3}=J^\mathrm{(M)}_{\mathrm{H},1\rightarrow4}.
	\label{}
	\end{equation}
We found the currents are identical everywhere when it operated under the steady state. This identical current is defined as $J$.
\section{Results}

To demonstrate the performance of this thermal phonon lasing, we run the simulation under two different cases: the phonon-free case and with lattice displacement field.  In the first case, the coupling between the middle four-level system and phonon field is neglected ( the coupling strength $g$ is set to be 0).  In phonon coupled case, we take into account the coupling to demonstrate the amplification of phonon field. Before that, we define an important physical property associated with lasing action: inversion. The total inversion is defined as $(\rho^\mathrm{(M)}_{22}-\rho^\mathrm{(M)}_{11})+(\rho^\mathrm{(M)}_{44}-\rho^\mathrm{(M)}_{33})$. This phonon laser is assumed to be operated under resonance condition ($\varepsilon^{\mathrm{(T)}}_{12}=\varepsilon^{\mathrm{(M)}}_{14}$ and $\varepsilon^{\mathrm{(B)}}_{12}=\varepsilon^{\mathrm{(M)}}_{23}$), which is shown as the optimized condition to achieve maximum inversion \cite{Pawel}.

In the simulation, parameters are set to be: $\varepsilon^{\mathrm{(T)}}_{12}=\varepsilon^{\mathrm{(M)}}_{14} = 30$ $\mathrm{meV}$, $\varepsilon^{\mathrm{(B)}}_{12}=\varepsilon^{\mathrm{(M)}}_{23}= 25$ $\mathrm{meV}$, $\lambda_\mathrm{ML}=\lambda_\mathrm{MR}= 0.03$ $\mathrm{meV}$, $\gamma_\mathrm{H}=\gamma_\mathrm{C}= 1 \mathrm{ps}^{-1}$, $\gamma_\mathrm{sys12}=\gamma_\mathrm{sys34}=\gamma_\mathrm{sys} $. The temperatures for the left reservoir (hot bath) and right reservoir (cold bath) are set to be $400\mathrm{K}$ and $100\mathrm{K}$ correspondingly.  

\begin{figure}
	\centering
	\includegraphics[width=8cm]{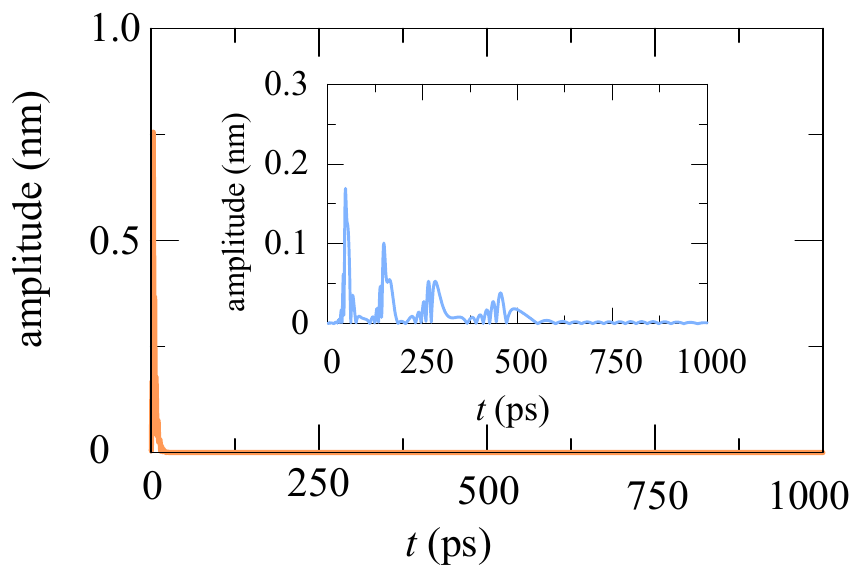}
	\caption{The envelope of lattice displacement fields, $\langle \hat{b}_1 \rangle$ (orange) and $\langle \hat{b}_2 \rangle$ (blue).}
	\label{fig:phonons}
\end{figure}

Then in the simulation phonon field presented, the parameters associated with phonons are set to be: $\Gamma= 1$ $\mathrm{ps}^{-1}$, $g= 2.25$  $\mathrm{ps}^{-1}$, and $u_0= 20$ $\mathrm{pm}$. We also change the some parameters to the following value: $\varepsilon^{\mathrm{(T)}}_{12}=\varepsilon^{\mathrm{(M)}}_{14} = 29$ $\mathrm{meV}$, $\varepsilon^{\mathrm{(B)}}_{12}=\varepsilon^{\mathrm{(M)}}_{23}= 25$ $\mathrm{meV}$, $\lambda_\mathrm{ML}=\lambda_\mathrm{MR}= 0.08$ $\mathrm{meV}$, $\gamma_\mathrm{H}=\gamma_\mathrm{C}= 5 \mathrm{ps}^{-1}$, $\gamma_\mathrm{sys12}=\gamma_\mathrm{sys34}=\gamma_\mathrm{sys} =0$. In Fig.~\ref{fig:phonons}, we present the expectation value of the phonon field operators $\langle \hat{b}_1 \rangle$ and $\langle \hat{b}_2 \rangle$  as a function of time. The initial values of $\langle \hat{b}_1 \rangle$ and $\langle \hat{b}_2 \rangle$ are set to be $10^{-3}$, which are caused by the vacuum fluctuation of the phonon field. The phonon field $\langle \hat{b}_1 \rangle$ and $\langle \hat{b}_2 \rangle$  increases dramatically at beginning, therefore shows its ability to amplify the phonon field. However, the damping is too strong, therefore the amplified field cannot last for long. We can observe only one pulse in $\langle \hat{b}_1 \rangle$, associated with the initial fast increasing of the population inversion between $|1\rangle$ and $|2\rangle$. The phonon field $\langle b_1 \rangle$ is amplified to the maximum simultaneous when the coupling is turned on. The maximum amplified phonon field is much larger than initial conditions. However, dynamics of $\langle \hat{b}_2 \rangle$ is totally different. There are several pulses in $\langle \hat{b}_2 \rangle$, associated with the several increasing of the inversion between $|3\rangle$ and $|4\rangle$. The amplitude of each pulse is twice smaller compared with $\langle \hat{b}_1 \rangle$.

\begin{figure}
	\centering
	\includegraphics[width=8cm]{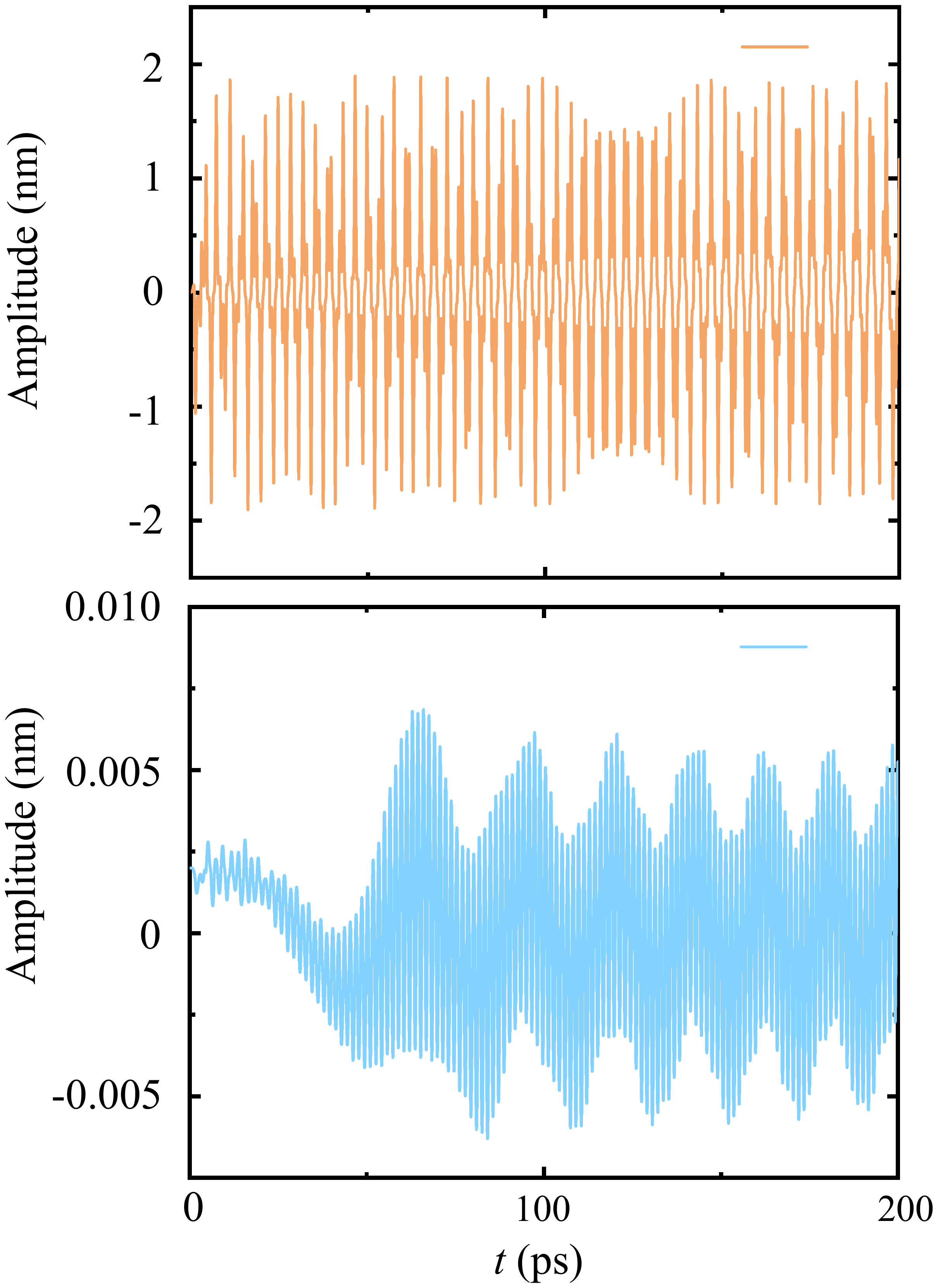}
	\caption{Lattice displacement field amplitude - full dependence.}
	\label{fig:Phonon Field_b1}
\end{figure}

\section{Thermodynamic Analysis}
According to the first law of thermodynamics (law of conservation of energy), the energy of the system changes when it have heat exchanges or power inpur or output. Therefore heat absorption from hot reservoir is related to the energy change in left subsystem, written as
\begin{equation}
	\dot{Q}_\mathrm{H}=\varepsilon^\mathrm{(T)}_{12}J^\mathrm{(T)}_{1\rightarrow2}
	\label{}
\end{equation}
This also holds for the heat releasing to the cold reservoir
\begin{equation}
	\dot{Q}_\mathrm{C}=\varepsilon^\mathrm{(B)}_{12}J^\mathrm{(B)}_{1\rightarrow2}
	\label{}
\end{equation}
The dissaption released heat
\begin{equation}
	\dot{Q}_{\mathrm{D},12}=\varepsilon^\mathrm{(M)}_{12}J^\mathrm{(M)}_{\mathrm{D},1\rightarrow2}
	\label{}
\end{equation}
\begin{equation}
	\dot{Q}_{\mathrm{D},34}=\varepsilon^\mathrm{(M)}_{34}J^\mathrm{(M)}_{\mathrm{D},4\rightarrow3}
	\label{}
\end{equation}
The heat exchanges accompany with the change of entropy. The total entropy production of the environment is \begin{equation}
	\dot{S}=-\frac{\dot{Q}_\mathrm{H}}{T_\mathrm{H}}+\frac{\dot{Q}_\mathrm{C}}{T_\mathrm{C}}+\frac{\dot{Q}_{\mathrm{D},12}}{T_\mathrm{sys}}+\frac{\dot{Q}_{\mathrm{D},34}}{T_\mathrm{sys}}\ge 0
	\label{Clausius Inequality}
\end{equation}
If there is no dissapator
\begin{equation}
	\dot{Q}_{\mathrm{D},34}=\dot{Q}_{\mathrm{D},12}=0
	\label{}
\end{equation}
When the phonon lasers is operated under steady-state, we have
\begin{equation}
	\frac{\varepsilon^\mathrm{(B)}_{12}}{\varepsilon^\mathrm{(T)}_{12}}\ge \frac{T_\mathrm{C}}{T_\mathrm{H}}
	\label{}
\end{equation}
The minimum temperature difference to drive this system can be determined by the Clausius Inequality (Eq.\ref{Clausius Inequality})
\begin{equation}
	\Delta T\ge (\frac{\varepsilon^\mathrm{(T)}_{12}}{\varepsilon^\mathrm{(B)}_{12}}-1)T_\mathrm{C}
	\label{}
\end{equation}
We define this minimum temperature as the thermal  threshold to achieve lasing action. Below this threshold, this device cannot generate any phonon output. The thermal threshold is found to be the linear function of the temperature of thermal reservoir and the phonon energy, which is consistent with previous numerical result \cite{Pawel}.

Analog to the heat engine, we can define the efficiency of our phonon laser
\begin{equation}
	\eta_{\mathrm{ideal}}=\frac{\dot{Q}_\mathrm{H}-\dot{Q}_\mathrm{C}}{\dot{Q}_\mathrm{H}}=1-\frac{\varepsilon^\mathrm{(B)}_{12}}{\varepsilon^\mathrm{(T)}_{12}}\le1- \frac{T_\mathrm{C}}{T_\mathrm{H}}=\eta_\mathrm{Carnot}
	\label{}
\end{equation}
The upper bound is found to be the Carnot Efficiency, which is expected when this phonon laser is treated as heat engine. While when the dissipation is consider and the laser is operated under irreversible condition, some additional term is added to reduce the efficiency.
\begin{equation}
	\eta=\frac{\dot{Q}_\mathrm{H}-\dot{Q}_\mathrm{C}-\dot{Q}_{\mathrm{D},12}-\dot{Q}_{\mathrm{D},34}}{\dot{Q}_\mathrm{H}}
	\label{}
\end{equation}

\begin{equation}
	\eta=1-\frac{T_\mathrm{C}}{T_\mathrm{H}}-\frac{T_\mathrm{C} \dot{S}}{Q_\mathrm{H}}-(1-\frac{T_\mathrm{C}}{T_\mathrm{sys}})\frac{(\dot{Q}_{\mathrm{D},12}+\dot{Q}_{\mathrm{D},12})}{Q_\mathrm{H}}\le\eta_\mathrm{Carnot}
	\label{}
\end{equation}
The partition of the dissipation and the total entropy production contribute to the reduction of the efficiency. The temperature of the middle system also have a influence on the efficiency.To increase the temperature of middle system and reduce the dissipation is able to optimize the efficiency.
	\section{Conclusion}
We proposed a four-level phonon laser medium composed of quantum subsystems. The explicit expression of Hamiltonian of this coupled quantum system have been presented to illustrate the coupling mechanism between energy filters and gain medium, and between the gain medium and phonon field. The system has been described using the density matrix formalism, of which the motion is determined by Quantum Master Equation within Linblad form. We have shown that this phonon lasing medium can achieve population inversion as well as amplification of lattice displacement field. Moreover, the diagonal terms of Quantum Master Equations have been rewritten by currents and simplified by the resonance assumption. The verification of this resonance assumption has been shown as the result of numerical methods we used. Based on the modified Quantum Master Equation, we conduct a thermodynamics analysis by defining the heat flow. We found that there is an upper bound of efficiency of this phonon laser concept, which is known as Carnot efficiency. 
\bibliographystyle{prsty}
\bibliography{reference}

\end{document}